**Radio-loud CMEs from the disk center lacking shocks at 1 AU**

N. Gopalswamy[1], P. Mäkelä[1,2], S. Akiyama[1,2], S. Yashiro[1,2], H. Xie[1,2], R. J. MacDowall[1], and M. L. Kaiser[1]

[1]NASA Goddard Space Flight Center, Greenbelt, MD, USA

[2]The Catholic University of America, Washington, DC, USA

A coronal mass ejection (CME) associated with a type II burst and originating close to the center of the solar disk typically results in a shock at Earth in 2-3 days and hence can be used to predict shock arrival at Earth. However, a significant fraction (about 28%) of such CMEs producing type II bursts were not associated with shocks at Earth. We examined a set of 21 type II bursts observed by the Wind/WAVES experiment at decameter-hectometric (DH) wavelengths that had CME sources very close to the disk center (within a central meridian distance of 30 degrees), but did not have a shock at Earth. We find that the near-Sun speeds of these CMEs average to ~644 km/s, only slightly higher than the average speed of CMEs associated with radio-quiet shocks. However, the fraction of halo CMEs is only ~30%, compared to 54% for the radio-quiet shocks and 91% for all radio-loud shocks. We conclude that the disk-center radio-loud CMEs with no shocks at 1 AU are generally of lower energy and they drive shocks only close to the Sun and dissipate before arriving at Earth. There is also evidence for other possible processes that lead to the lack of shock at 1 AU: (i) overtaking CME shocks merge and one observes a single shock at Earth, and (ii) deflection by nearby coronal holes can push the shocks away from the Sun-Earth line, such that Earth misses these shocks. The probability of observing a shock at 1 AU increases rapidly above 60% when the CME speed exceeds 1000 km/s and when the type II bursts propagate to frequencies below 1 MHz.

## 1. Introduction

Type II radio bursts are excellent indicators of shocks near the Sun and in the interplanetary (IP) medium. The longer the wavelength range of the type II radio bursts, the greater is the distance traveled by the underlying shock [*Gopalswamy et al.*, 2005]. Coronal mass ejections (CMEs) are the drivers behind these shocks. CMEs associated with type II bursts (radio-loud CMEs) are highly likely to result in shocks at Earth: it was recently found that 66% of IP shocks had type II burst association, while 34% were not [*Gopalswamy et al.*, 2008a,b; 2010a]. However, a large number of CMEs with type II bursts do not have shocks at Earth. It is understandable that CMEs originating near the limb may not have a shock signature at Earth because of the limited longitudinal extent of the shock. What is surprising is that there are many radio-loud CMEs originating from close to the disk center, yet they produce no shock signature at Earth. Given the fact that even some radio-quiet CMEs produce a shock signature at Earth [*Gopalswamy et al.*, 2010a], it is important to understand why shocks near the Sun inferred from type II radio bursts are not seen at Earth. Clarification of this issue is very important in using the type II bursts as early indicators of space weather events.

In this work, we investigate the set of type II bursts at decameter-hectometric (DH) wavelengths produced by CMEs originating from close to the disk center (central meridian distance (CMD) $\leq 30^{\circ}$) and examine the source properties to understand why the associated shocks are not observed at Earth.

## 2. Description of the Data Sets

The starting point of this investigation is the set of all type II bursts detected by the Radio and Plasma Wave (WAVES) experiment [*Bougeret et al.*, 1995] on board the Wind spacecraft. These type II bursts and the associated CMEs detected by the Large Angle and Spectrometric Coronagraph [LASCO, *Brueckner et al.*, 1995] on board the Solar and Heliospheric Observatory (SOHO) mission are cataloged and made available on line at the CDAW Data Center (http://cdaw.gsfc.nasa.gov/CME_list/radio/waves_type2.html; Yashiro et al., 2004; Gopalswamy et al., 2009a). The field of view of LASCO corresponds to the spatial domain in which the WAVES experiment observes the type II radio bursts, so pairing the type II bursts with the CMEs that produce them is rather straightforward. There are 342 type II bursts observed in the decameter-hectometric (DH) wavelength domain sampled by the WAVES experiment during solar cycle 23 (1996 – 2008, inclusive). The last type II burst observed was on 2008 April 26 [*Gopalswamy et al.*, 2009b]. The solar source location of CMEs that produced the type II bursts is defined as the heliographic coordinates of the associated eruption: H-alpha flare location if available from the Solar Geophysical Data (SGD) or by playing movies of the EUV images obtained by the Extreme-ultraviolet Imaging Telescope [EIT, *Delaboudinière et al.* 1995] to identify the location of an associated disk activity such as EUV dimming or post eruption arcades [see *Gopalswamy et al.*, 2007; 2009c for details on CME source identification]. The solar source locations are also listed in the type II burst catalog.

Over the same period, *Gopalswamy et al.* [2010a] reported 222 IP shocks with the associated CMEs. The shocks were detected in situ by one or more of the three spacecraft at L1: SOHO, Wind and the Advanced Composition Explorer (ACE). These shocks, along with their solar source properties, are listed in *Gopalswamy et al.* [2010a] as an electronic supplement. The shocks were compiled from the lists available at the web sites of observing spacecraft: Wind (http://lepmfi.gsfc.nasa.gov/mfi/ip_shock.html), ACE (http://www-ssg.sr.unh.edu/mag/ace/ACElists/obs_list.html#shocks), and SOHO

(http://umtof.umd.edu/pm/FIGS.HTML). After carefully eliminating shocks associated with corotating interaction regions (CIRs), we arrived at the list of 222 shocks, all of which had overlapping coronagraphic observations.

Out of these 222 shocks, 145 were radio loud, 76 were radio quiet, and for one shock, there was no radio data. The shock list is available on line along with the information on the associated CMEs, flares and type II radio bursts (http://iopscience.iop.org/0004-637X/710/2/1111/fulltext/apj_710_2_1111.tables.html). Instead of starting from IP shocks, if we start with DH type II bursts, we found that the majority of the DH type II bursts were not associated with the shocks listed in *Gopalswamy et al*. [2010a]: 206 out of the 342 DH type II bursts (or ~60%) were not associated with a shock at Earth. The focus of the present study is the 206 DH type II bursts that did not have a shock at Earth. The driving CMEs (from SOHO/LASCO) were readily identified for these type II bursts, but 59 of these CMEs were back-sided (DH type II bursts from many back-sided CMEs are observed because the low-frequency radio sources are not occulted). Two events on 2001 March 27 listed in the type II burst catalog have been dropped because the events were so short that their spectral shape is not clear enough to be called as type II bursts. The remaining 145 DH type II bursts without a shock at Earth had their CMEs originating from the frontside of the Sun (CMD <90$^o$). We refer to the CMEs as RLNS CMEs (Radio Loud No Shock at Earth). The source longitude distribution and heliographic coordinates of the 145 RLNS CMEs are shown in Fig. 1. The longitude distribution shows that the number of RLNS CMEs is generally low near the disk center compared to those at the limbs. The limb RLNS CMEs are understandable from geometrical considerations: limb CMEs may be driving a shock, but the shock may not be extended sufficiently to reach to the Sun-Earth line. The problem is why the disk-center CMEs that have shock signatures near the Sun (DH type II burst) do not have shock signatures at Earth. Therefore, we consider a subset 21 RLNS CMEs that originate within a central meridian distance (CMD) of 30° (see the events within the circled region in Fig. 1). It is important to understand these events because the lack of a shock at 1 AU is contradictory to the expectation.

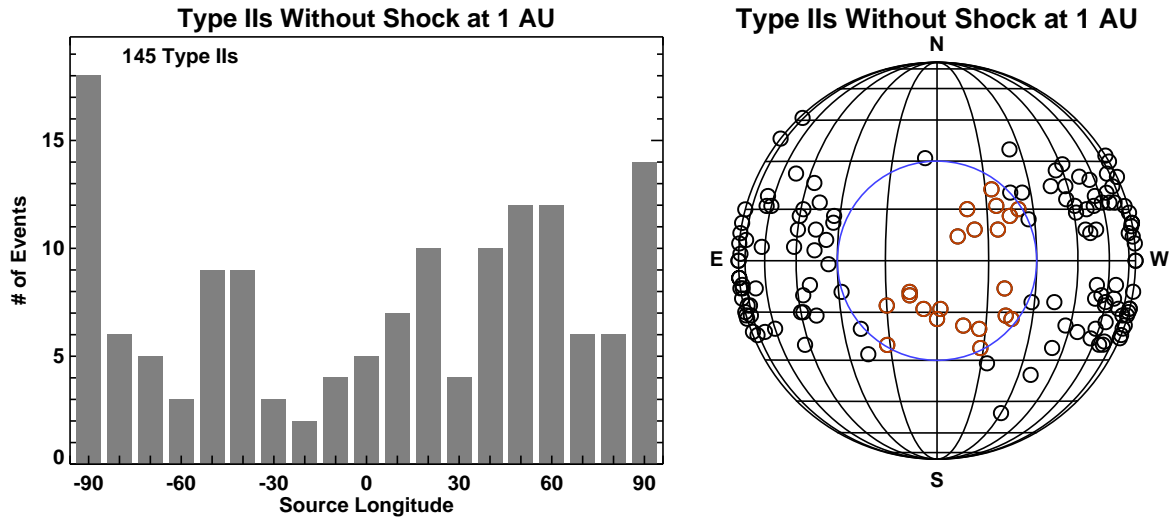

*Figure. 1 (a) The longitudinal distribution of the RLNS CMEs. (b) Heliographic coordinates of RLNS CMEs, with the disk-center sources (CMD ≤30°) shown encircled. The central CMEs are the subject of this investigation, because the shocks from these CMEs are expected to arrive at Earth. We have excluded 59 behind-the-limb CMEs.*

The latitude distribution and the time variation of the latitudes of the frontside RLNS CMEs are shown in Fig.2. The CME latitudes are generally distributed within ±30° indicating that most of the CMEs originate from the active region belt. RL CMEs are generally more energetic and hence are produced in the active region belt. This is further confirmed by the plot of CME latitudes as a function of time, which resembles the sunspot butterfly diagram. The disk-center RLNS CMEs have a similar latitude variation with time, indicating more such CMEs during the maximum phase compared to the rise and declining phases.

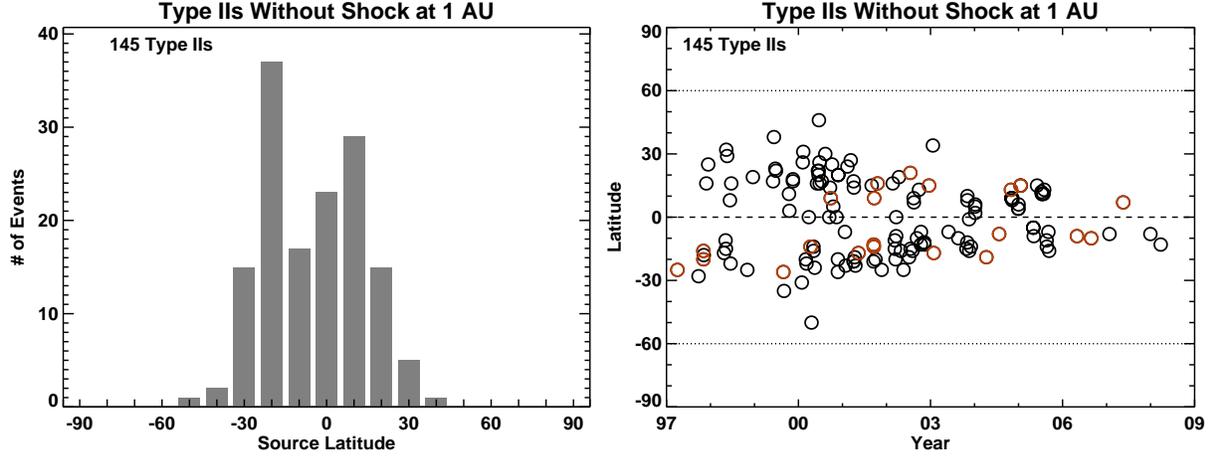

*Figure 2. Latitude distribution (a) and time evolution (b) of the solar sources of the RLNS CMEs. The red circles are the disk-center events. The solar sources are confined to the active region belt and follows the sunspot butterfly diagram.*

For comparison, we use the 136 RL CMEs (342 minus 206) that had a shock at Earth. We refer to these CMEs as RLS (radio-loud with shock at Earth). We have a smaller number of RLS events here because the radio loudness is based on DH type II bursts alone. In *Gopalswamy et al.* [2010a] events associated with purely metric type II bursts or purely kilometric type II bursts were also considered as radio loud. There was either CME data gap or the CMEs were too faint in 15 cases. Two of these RLS CMEs originated from behind the limb, so we do not know their heliographic coordinates. Thus the remaining 119 RLS CMEs are frontsided. In order to compare with the disk-center RLNS events, we consider a subset of 53 RLS events occurring with CMD $\leq 30°$. In other words, there are 74 disk-center radio-loud CMEs, of which 21 (or 28%) do not have a shock at Earth. Finally, we consider a set of 39 CMEs orignating close to the disk center (CMD $\leq 30°$) that are radio quiet, but associated with shocks at Earth (RQS CMEs). The RQS CMEs are a subset of the 76 RQ CMEs reported in Table 1 of *Gopalswamy et al.*

[2010a]. In summary, the RLS and RQS CMEs have an associated shock at Earth, while the CMEs in the RLNS population have no shock association. We would like to find out why.

Table 1 lists the properties of the 21 type II bursts belonging to the disk-center RLNS population (CMD≤ 30º). The first two coulumns give the start and end times of the type II burst as observed in the Wind/WAVES dynamic spectrum. The third and fourth columns give the starting (F1) and ending (F2) frequencies of the type II burst in MHz. The highest frequency of the WAVES experiment is 14 MHz and the lowest frequency is 20 kHz. The presence of a metric type II (mII) component is indicated in the fifth column (Y = yes; N = no). The heliographic coordinates of the solar sources of the CMEs are given in column 6, with the corresponding NOAA active region (AR) number listed in column 7. All the CMEs originated from numbered ARs, except for one associated with a filament eruption (noted as FILA in column 7). The soft X-ray flare importance is given in column 8. The first appearance time, the central position angle (CPA), the angular width in the sky plane (W in degrees) and the sky-plane speed (V in km/s) of the CMEs are given columns 9 -12. Column 13 indicates whether the CME was associated with an SEP event (Y = yes; N = no; HiB = high background). The last column gives the coronal hole influence parameter (CHIP), which is taken as an indicator that the CME might have been deflected from its trajectory by nearby coronal holes. CHIP is the magnitude of a fictitious force directly proportional to the area of the coronal hole and the average magntic field in the coronal hole, while inversely proportional to the square of the dustance between the coronal hole centroid and the eruption region [see *Gopalswamy et al*. 2009c for more details].

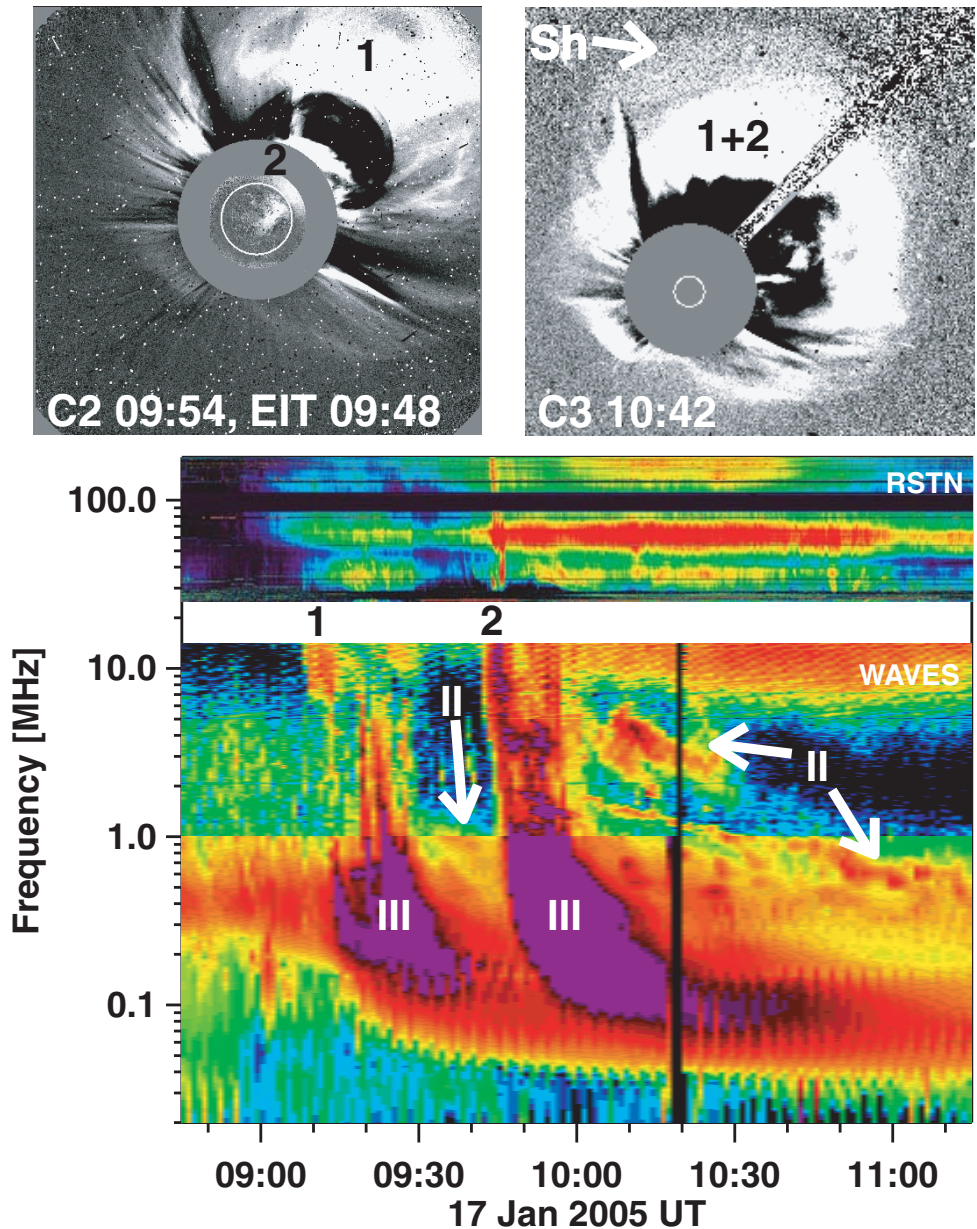

*Figure 3. (top) The 2005 January 17 CME at 09:30 UT (CME1) being overtaken by a second CME at 09:54 UT (CME2) resulting in a compound CME by 10:42 UT. Only a single shock (Sh) was seen at this time. (bottom) Composite radio dynamic spectrum constructed using observations from the Radio Solar Telescope Network (RSTN) and the Radio and Plasma Wave Experiment (WAVES) on board the Wind spacecraft showing type III, type II and type IV bursts associated with CME1 and CME2.*

Among the 21 RLNS CMEs listed in Table 1, the 2005 January 17 event is exceptional with a high speed of 2094 km/s and originating from the super active region 10720 located at N15W25. The CME in this event was overtaken by another CME from the same region that was even faster (2547 km/s) and was launched just 37 min after the first one [*Gopalswamy et al.* 2006]. Figure 3 shows the two CMEs in the LASCO/C2 frame at 09:54 UT: part of the first CME (CME1) has already left the LASCO/C2 FOV and the second CME (CME2) is at a heliocentric distance of 3.02 Rs. The leading edge of CME1 can be extrapolated from the 09:42 UT height of 7.39 Rs at 09:42 UT to be 9.6 Rs, indicating a spatial sepration ~6.6 Rs between the leading edges of the two CMEs at 09:54 UT. The two CMEs cannot be distinguished in the 10:42 UT LASCO/C3 frame because they have merged and one can see the diffuse shock sheath ahead of the resultant CME. Figure 3 also shows the Wind/WAVES radio dynamic spectrum, in which the two eruptions can be distinctly seen as two sets of type III, type II and type IV bursts. The type II emission from CME1 is fragmented near 14 MHz, but a distinct broadband component starts around 2 MHz at 09:30 UT and drifts to 0.4 MHz when the onset of the type III bursts due to the second eruption at 09:50 UT the type II burst. All the radio bursts associated with CME2 (type III, type II, and type IV) are more intense than those from CME1. Beyond the onset of CME2 radio bursts, no signatures of the first type II are seen. Since the two CMEs are from the same region and CME2 is faster, we think the shock from CME1 producing the first type II burst merged with the second shock by 10:42 UT as seen in the white-light image in Fig.3 at that time. There was also a single shock observed at 1 AU due to the combined CMEs. Thus, the 2005 Jan 17 event is a special case in that CME interaction is responsible for the lack of shock at Earth associated with the CME. Such interactions are not uncommon. A similar event was reported earlier in *Gopalswamy* [2004]. Note that CMEs can also be deflected due to collisions with other

CMEs [*Gopalswamy et al.* 2001a], but we do not have such events in the list in Table 1. In the further analysis of the disk-center events lacking a shock at Earth, we shall exclude this event from the statistical analysis to understand why so many RL CMEs did not have in-situ shock association.

3. Analysis and Results

While investigating the characteristic difference between RL and RQ shocks detected at 1 AU, *Gopalswamy et al.* [2010a] found that CME properties near the Sun were the most distinct. The RLS CME speeds were much higher and most of them were decelerating in the coronagraphic field of view. On the other hand, the RQS CME speeds were only slightly above the average speed of the general population and most of the CMEs were accelerating. The fraction of halos was also smaller in the RQS population compared to that in the RLS population. The difference in the shock properties at 1 AU was not significant because the two CME populations tend to attain the solar wind speed at large distances from the Sun due to momentum exchange with the solar wind [see e.g., *Gopalswamy et al.* 2000]. In this section we compare the properties of the RLS and RQS populations with those of the RLNS population.

3.1 CME Speed, width, and acceleration

Figure 4 compares the sky-plane speeds of the RQS, RLS, and RLNS CMEs whose solar sources have CMD $\leq 30^o$. Since all the three populations are from the same CMD range, they are subject to similar projection effects, so comparing the sky-plane speeds is justified. The RLS population stands out in terms of the CME speed: the mean speed (1281 km/s) is about twice as large as those of the other two populations. The range of speeds of the RLS population is also much larger: 300 to 2900 km/s compared to 100 to 900 km/s for the RQS and 300 to 1100 km/s for the RLNS populations. The three populations overlap significantly in the low-speed ends of the distributions. The distribution of RLNS population is somewhat similar to that of the RQS population, except that the average speed is slightly higher (644 km/s vs. 509 km/s). The speed bin with the maximum number of events is the same for RQS and RLNS. There is an extra bin at the high-speed end for the RLNS population and at the low-speed end for the RQS population. Thus based on the speed distributions, we can say that the RLNS population is closer to the RQS population, than to the RLS population.

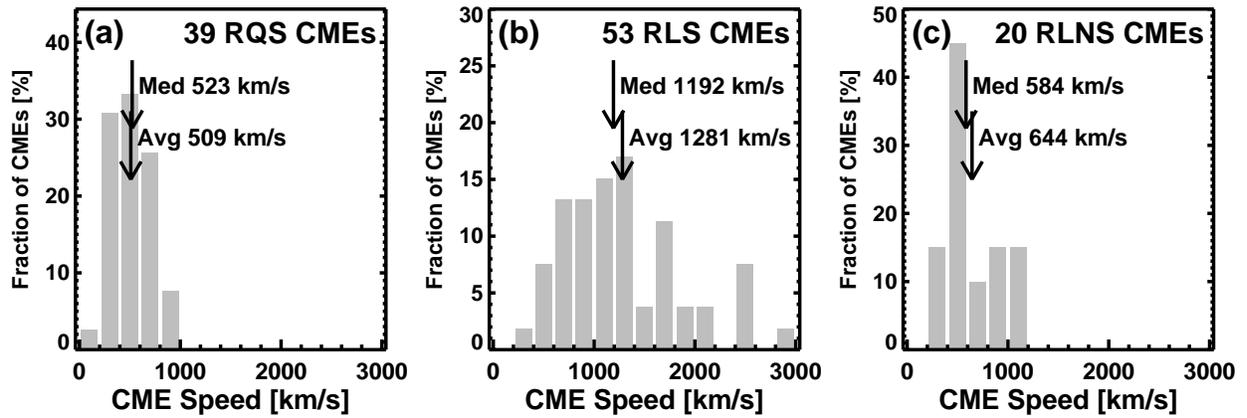

*Figure 4. The sky-plane speed distributions of the three populations: (a) radio quite CMEs with shock signatures at Earth (RQS), (b) radio-loud CMEs with shock signatures at Earth (RLS), and (c) radio-loud CMEs with no shock signatures at Earth (RLNS). The mean and median values of the distributions are marked on the plots.*

The CME width distributions for the three populations are shown in Fig. 5. One of the prominent features of the distributions is that the width = $360^o$ bin has the largest number of CMEs for all the three populations. Even the avearge width of non-halos is greater than $120^o$ in all three cases. In the case of RLS population, most CMEs are halos (91%), whereas slightly more than half (54%) of the RQS CMEs are halos. The lowest fraction of halos is in the RLNS population (only 30%). The fraction of halo CMEs in a CME population has been suggested as a measure of the average energy of the population [*Gopalswamy et al.*, 2007; 2010b]. Therefore, the RLNS CMEs seem to be less energetic compared to the other two populations. This result raises two possibilities. First, many of the non-halo CMEs in the RLNS population might have been ejected nonrdially because they do not appear as halos, even though their solar sources are close to the disk center. Second, many of these CMEs may have relatively smaller spatial extent that they were missed by the spacecraft making in situ measurements. More discussion on this issue can be found later.

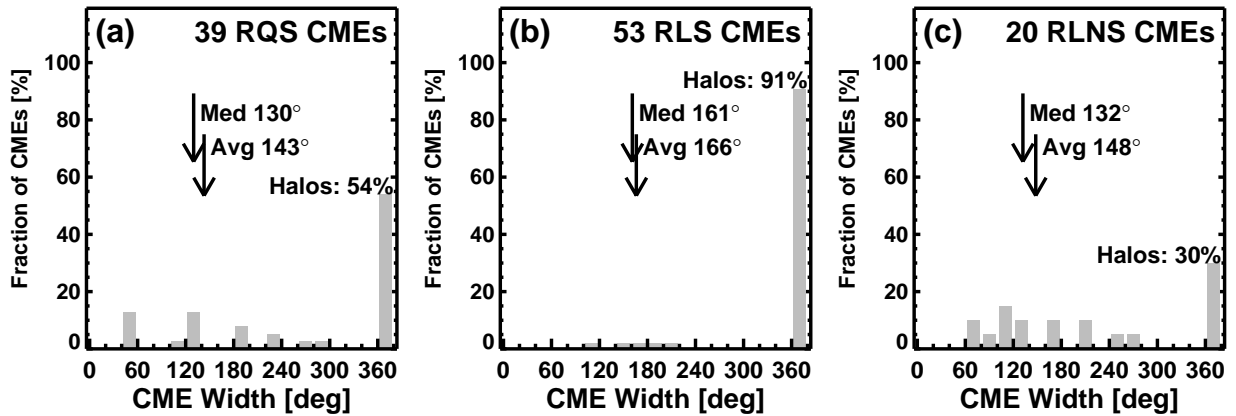

*Figure 5. The sky-plane widths of the three CME populations: (a) RQS, (b) RLS, and (c) RLNS. The mean and median values of the distributions are marked on the plots. The bin width is $20^o$. Width = $360^o$ corresponds to halo CMEs (the last bin in each histogram).*

One of the key differences between RQS and RLS CMEs was found to be their distinctly different residual acceleration within the coronagraphic field of view. Most of the RLS CMEs showed deceleration, while the RQS CMEs showed acceleration [*Gopalswamy et al.*, 2010a]. The RLNS CMEs do not fall into these extremes: accelerating and decelerating CMEs are roughly equal in number (9 vs. 11). Figure 6 shows the distribution of residual acceleration for the three populations. Deceleration dominates in the RLS population; acceleration dominates in the RQS population; the RLNS population is intermediate with roughly equal number of accelerating and decelerating CMEs.

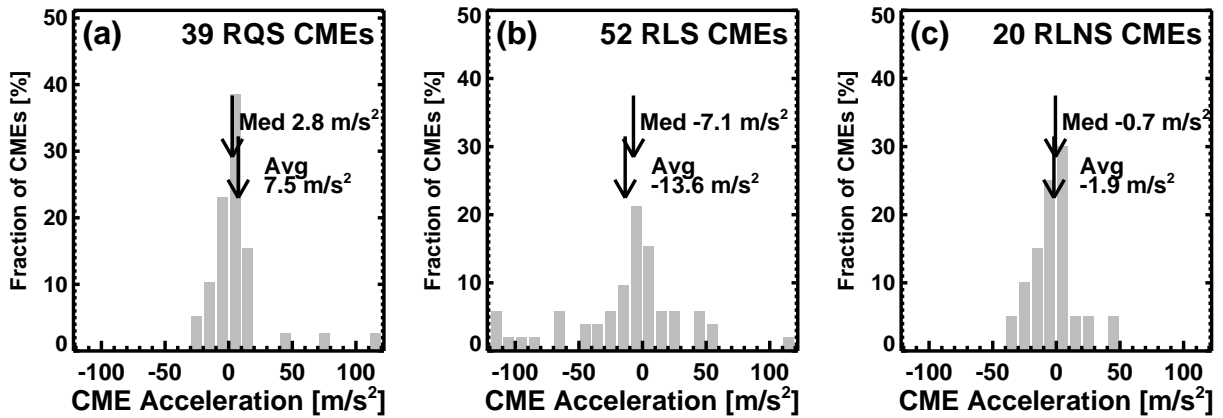

Figure 6. The sky-plane accelerations of the three CME populations: (a) RQS, (b) RLS, and (c) RLNS. The mean and median values of the distributions are marked on the plots.

### 3.2 Type II radio burst association

The selections of RLNS events in Table 1 is based on their association with DH type II bursts. However, to get the real starting frequency of the type II activity, we need to examine the metric type II association also. Information on the the metric type II bursts is available on line from the National Geophysical Data Center (NGDC):

ftp://ftp.ngdc.noaa.gov/STP/SOLAR_DATA/SOLAR_RADIO/SPECTRAL/Type_II/Type_II_1994-2009. Dynamic spectra from indiviual observatories were also examined to confirm the presence or absence of metric type II burst association. Among the 53 RLS CMEs, 14 (or 26%) did not have metric counterparts. Among the 20 RLNS CMEs roughly a similar fraction did not have a metric type II (5 out of 20 or 25%). For the remaining events, we identified the maximum frequency from the metric type II observtions and the minimum frequency from the Wind/WAVES dynamic spectrum. No attempt was made to separate the fundamental and harmonic components because this information is not available for all the events. The RLNS population had their starting frequency in the range 270 MHz to 3 MHz. The ending frequencies were mostly above 1 MHz, except for one event: the 2006 April 30 event had 0.2 MHz. In the 2006 April 30 event, the type II was observed in the metric domain (180 – 25 MHz) during 09:21 to 09:27 UT. A type II burst was also reported appear in the WAVES dynamic spectrum starting around 1 MHz at 11:30 UT. A close examination of the dynamic spectrum reveals that the starting frequency is more like 0.5 MHz, so this is strictly not a DH type II burst. Thus barring the 2006 April 30 event, the type II bursts in the RLNS population ended in the RAD2 spectral domain (14 -1 MHz). The type II burst starting frequencies in the RLS population are similar to those in the RLNS population, but the ending frequencies are markedly different. The ending frequencies are below 1 MHz in 39 out of the 53 events (or 74%). From the low ending frequencies one can infer that the CME-driven shocks continued to be strong in the IP medium and accelerated electrons to produce the type II radio bursts. As for the RLNS events, one can conclude that the shocks weakened considerably before the driving CMEs left the coronagraphic field of view.

Figure 7 compares the starting and ending frequencies of the RLNS populations with those of the RLS population over the frequency range corresponding to the local plasma frequency at the Wind spacecraft (~20 kHz) to metric frequencies. Clearly, the maximum number of events can be found in the bin with values in the range 2.0 to 2.5, corresponding to a central frequency of 178 MHz. This is the typical starting frequency of metric type II bursts and corresponds to the coronal region of lowest Alfven speed [*Gopalswamy et al.*, 2001b]. On the other hand, the ending frequencies clearly peak at drastically different frequencies: the maximum number of events can be found in the bin 0 to 0.5 (RLNS) and -1.5 to -1.0 (RLS). These bins roughly correspond to 1.8 MHz (RLNS) and 0.06 MHz **(RLS)**. The higher ending frequencies in the RLNS population is consistent with the lower energy of the CMEs inferred from the smaller avarage speed and the smaller fraction of halos, supporting the the hierarchical relationship between the spectral range of type II bursts and the energy of the associated CMEs [*Gopalswamy et al.*, 2005].

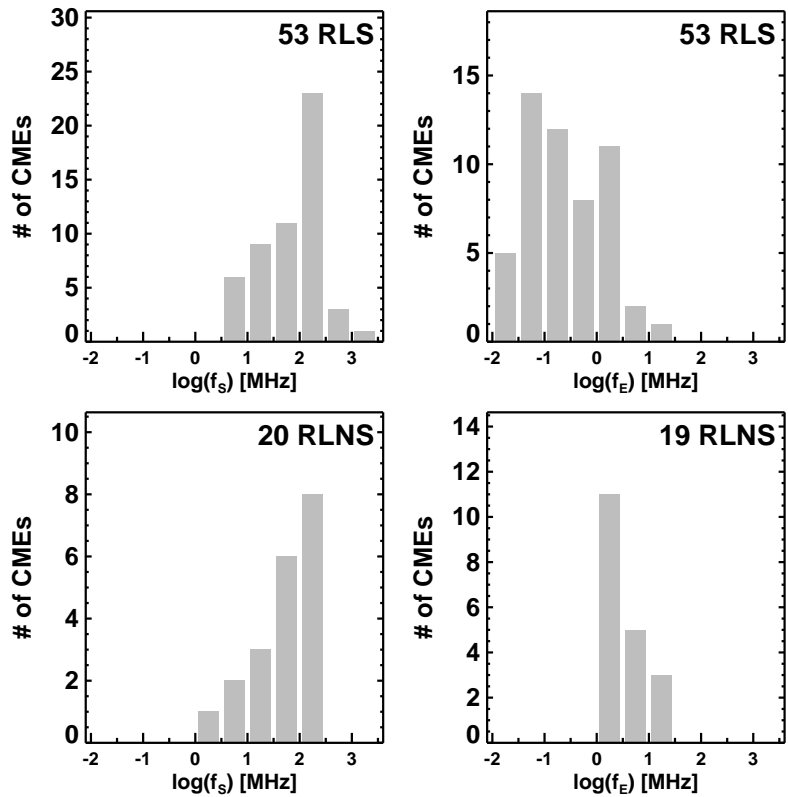

*Figure 7. The starting and ending frequency distributions of type II bursts associated with RLS and RLNS CMEs. The frequencies are in log scale. The bin size is 0.5. The bin labels shown correspond to the lower end of the bin. For example, the bin 0.2 represents the range 0.2 to 0.6 (1.58 to 3.98 MHz). In the left panels, the bins >0.6 represent metric type II bursts. In the right*

*panels, the highest bin corresponds to the high-frequency end of the Wind/WAVES spectral range (1.14 in log units or 14 MHz). The two events in the lower right are questionable (see text), indicating a sharp cutoff at 1 MHz for the ending frequencies.*

### 3.3 SEP Association

Large SEP events are thought to be indicators of powerful CME-driven shocks near the Sun. Recent investigations indicate that SEPs are released immediately after the start of the metric type II radio bursts, when the CMEs reach a height of ~1.5 Rs [see e.g., *Gopalswamy et al.*, 2012]. Since the RLNS CMEs are associated with type II bursts near the Sun, one may wonder if these CMEs are associated with SEP events. We searched in the GOES and SOHO/ERNE data for possible SEP events for a period of several hours following the CME onset. When we examined the composite plots that compare the GOES SEP intensity, CME height – time history, and the GOES soft X-ray flare, only three events were clearly seen above the background. The GOES data in Fig. 8 shows the particle enhancement in the >10 MeV and > 50 MeV channels for the 2001 December 19 RLNS event. The SEP intensity peaked at ~4 pfu (particle flux unit; 1pfu = 1 particle per $cm^2$.s.sr), which is considered as a minor event [*Gopalswamy et al.*, 2002]. Note that this CME is the fastest event in Table 1. The other two GOES SEP events are also minor events: a 7-pfu SEP event associated with the 2001 October 19 RLNS CME and a 1-pfu event on 2004 October 30. These two CMEs had only moderate speed (558 and 422 km/s, respectively).

We also examined the SOHO/ERNE data in five energy channels: 1.78 – 3.29 MeV, 3.29 – 6.42 MeV, 6.42 – 12.7 MeV, 13.8 – 28.0 MeV, and 25.9 – 50.8 MeV for all the RLNS events. Figure 9 shows the 1997 April 1 event, which has no obvious SEP enhancement in the GOES particle data. The ERNE data shows a clear SEP event in all the energy channels. The onset shows also velocity dispersion, with the onset being the earliest at the highest energy channel. The event lasts for about two days. A similar examination of the ERNE data around the time of the type II radio burst revealed that 10 of the 20 RLNS events (or 50%) had clear SEP events. For four events, no significant particle enhancement was found. In the remaining 6 events, the particle background was high due to preceding events, so it was not possible to say if there was an SEP event or not.

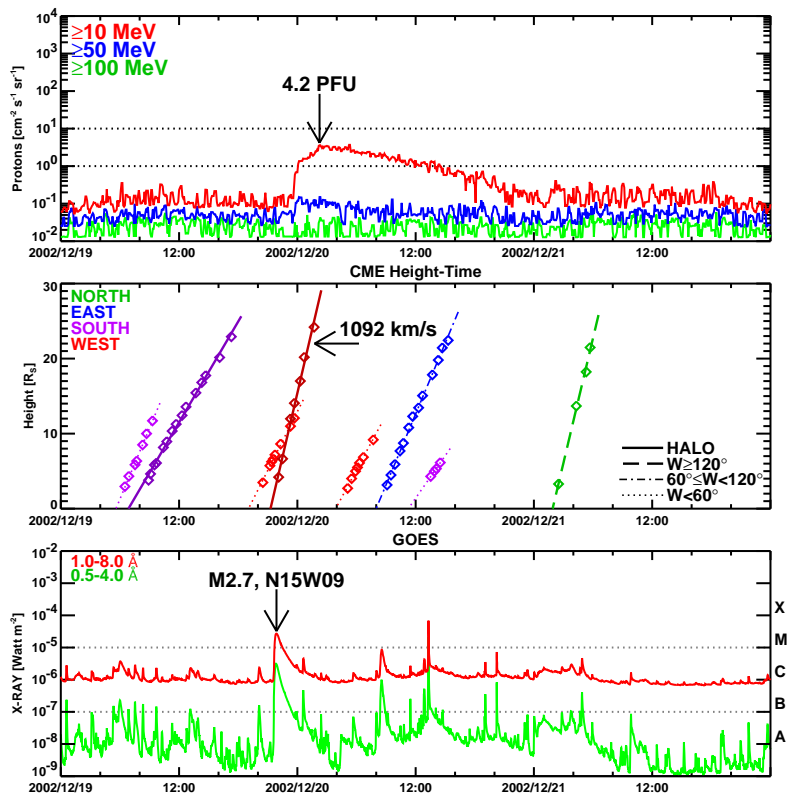

*Figure 8. (top) GOES proton intensity in three energy channels (>10 MeV, > 50 MeV, and > 100 MeV) as a function of time over a three-day interval starting from 2001 December 19. A minor SEP event can be seen starting towards the end of 2001 December 19 and lasting for slightly more than a day. (middle) height-time plots of CMEs over the same time interval. (bottom) the GOES soft X-ray light curve in two energy channels (1.0- 8.0 Å and 0.5 to 4.0 Å)*

*with the heliographic coordinates of the flares marked. This is the fastest CME in the RLNS population associated with an M2.7 flare originating from close to the disk center (N15W09).*

A similar analysis of the RQS events reveals three significant GOES events: a large SEP enhancement (>10 pfu) superposed on the decaying part of a preceding event that started on 2001 April 18 and two minor events. Although the large enhancement seems to be associated with a moderate CME (663 km/s in the LASCO FOV) on 2001 April 19 at 12:30 UT and an M-class flare from the northwest quadrant of the Sun, one cannot rule out the possibility of interplanetary modulation due to a pressure pulse (clearly seen in the WAVES/TNR data, not shown) associated with the 2001 April 18 CME. The minor SEP event on starting on 2001 March 25 is very complex, probably contributed by two CMEs (2001 March 24 at 20:50 UT and March 25 at 17:06 UT) and two in-situ shock enhancements (associated with shocks at 2 and 18 UT on March 27). The 2004 July 22 event also seems to be an energetic storm particle (ESP) event associated with the interplanetary shock at 09:45 UT due to the 2001 July 20 CME at 13:31 UT. There were five other events with enhancements in the ERNE data. Three of these events seem to have possible type II association at frequencies below 1 MHz masked by ongoing noise storms. The remaining two events with definite SEP enhancement need further investigation. Thus we conclude that the RQS population has almost no SEP association.

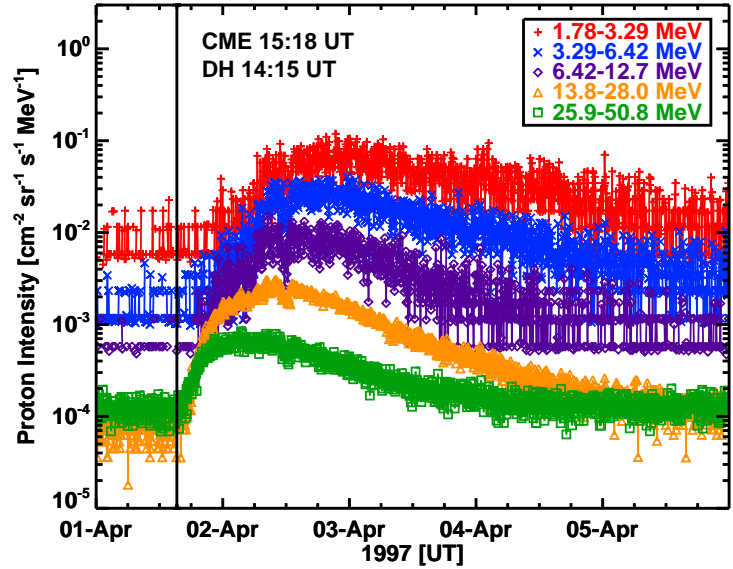

*Figure 9. SEP intensity recorded by SOHO/ERNE in five energy channels showing that the 1997 April 1 RLNS event was associated with a small particle event that lasted for more than a day.*

The RLS population had the highest SEP association. By examining the GOES particle intensities, we found that there were 23 large SEP events and 14 minor SEP events in the RLS population. When we examined the ERNE plots for the remaining events, we found that many CMEs occurred when the background SEP level was high. Enhancements were found on top of the high background immediately after the CME in several cases, but or 3 events it was not possible to say if there was a new enhancement above the elevated background. No enhancement was observed only in the remaining 13 events. We can conclude that at least 37 of the 50 RLS events (or 74%) had SEP association (excluding the three high-background events). Thus the SEP association rate is the highest (74%) for the RLS events followed by RLNS events (50%) and the least for RQS events (15%). This result demonstrates that the presence of a shock near the Sun is more important for the production of SEPs than its presence at 1 AU and consistent with the importance of type II burst association [Cliver et al., 2004]. In a previous study, *Mäkelä* *et al.* [2011] investigated the particle enhancement at the shock at 1 AU and found that the RLS events are more likely to be associated with ESP events than the RQS events.

## 3. 4. Flare Association

Table 2 shows the number of X, M, C, and B-class flares in the RQS, RLS, and RLNS populations. The RLS events are associated with the strongest flares (81% X and M class) followed by the RLNS events (60% X and M class), while the RQS events are associated with the weakest flares (mostly C-class or lower; 16% M-class and no X-class flares). The results are consistent with the flare – type II relationship discussed in *Gopalswamy* [2006]. Larger flares are associated with more energetic CMEs. The RLS events are associated with type II bursts occurring over the widest frequency range and hence are associated with the strongest flares. The RLNS events are associated with type II bursts occurring only in the near-Sun IP medium, so the associated flare sizes are intermediate. The RQS events have no type II burst at all and the flare sizes are the smallest.

## 5. Discussion and Conclusions

Type II radio bursts, especially at frequencies below 14 MHz, have been used as an indicator of powerful CMEs that drive shocks in the IP medium. This is useful information because the same shocks accelerate SEPs that have important space weather effects. The shocks may also arrive at Earth causing a sudden commencement and/or an energetic storm particle event. The results presented in this paper tell us tell us that not all shocks observed near the Sun and reach 1 AU contrary to the expectation. We examined various reasons as to why the shocks do not propagate far into the IP medium.

The primary finding of this paper is that a significant fraction (21 out of 74 or 28%) of shocks near the Sun, identified based on their ability to produce type II radio bursts, do not arrive at Earth even though the driving CMEs originate close to the disk center and hence are expected to arrive at Earth. In order to understand why these shocks are not observed at 1 AU, we examined the kinematic properties of the driving CMEs, the spectral range of type II bursts, the SEP association rate, and the flare properties. We then compared these properties with those of two other CME populations: (i) the radio-loud CMEs that did have their shocks arriving at 1 AU (the RLS population), and (ii) the radio-quiet CMEs that did not have type II bursts but did have their shocks arriving at 1 AU (RQS population).

Examination of the kinematic properties reveals that the RLNS CMEs are the slowest among CMEs that are associated with DH type II bursts. The average and median speeds of RLNS CMEs are a factor of 2 smaller than the corresponding speeds of the RLS CMEs. On the other hand, these speeds are similar for RLNS and RQS CMEs. Interestingly, the speeds are also similar to CMEs that produce type II bursts over limited spectral ranges (purely metric type II bursts and purely kilometric type II bursts, see *Gopalswamy et al.* [2010a]. The CME widths as indicated by the fraction of halos are also smaller for the RLNS CMEs. The halo fraction is only 30% for the RLNS CMEs, compared to 54% for the RQS and 91% for the RLS CMEs. The halo fraction is thus distinctly different for the RLNS CMEs. The smaller speed and halo fraction of the RLNS CMEs is a clear indication that the RLNS CMEs have low kinetic energy to start with. This has two consequences: (i) the CMEs may not have enough energy to drive shocks to large distances and (ii) the narrower width means the shock can be missed at Earth if the ejection is inherently non-radial or the CME is deflected by nearby coronal holes. The third kinematic property we considered is the acceleration of CMEs within the coronagraphic field of view. Note that this is not the initial acceleration, which could be substantially large (often more than 1 km/s$^2$). The RLS CMEs generally decelerate, while the RQS CMEs generally accelerate. *Gopalswamy et al.* [2010a] explained that the continued acceleration of RQS CMEs was responsible for their shock-driving capability at large distances from the Sun. The RLNS CMEs do not show any trend, but closer to the RLS CMEs in that they have a small negative acceleration, consistent with the shock formation closer to the Sun. The kinematic properties of CMEs in the three populations are consistent with the known correlation between CME kinetic energy and the soft X-ray flare size. The RLS CMEs are associated with the largest flares, the RLNS CMEs with the intermediate flare sizes, and the RQS CMEs with the weakest flares.

The spectral range of type II bursts is another indicator of the CME energy. *Gopalswamy et al.* [2005] showed that type II bursts with emission components at all wavelength ranges (metric, DH, and kilometric) are associated with the most energetic CMEs. Type II bursts occurring only in the metric domain or in the kilometric domain have the lowest energy. The RLNS CMEs reported here have properties very similar to CMEs associated with the metric-only type II bursts [see *Gopalswamy et al.* 2005; 2010a]. The only difference is that the type II bursts in the RLNS population extend slightly beyond the metric domain and extend to the DH domain. The ending frequencies type II bursts produced by the RLNS CMEs were generally above ~1 MHz, which means the CMEs became radio quiet at distances of ~10 Rs. On the other hand, the ending frequencies of type II bursts produced by the RLS CMEs were much lower – 60 kHz (in the kilometric domain). Lack of type II bursts at heliocentric distances of ~10 Rs implies that either the shock is weak and unable to accelerate sufficient number of electrons or it has decayed. On the other hand, type II bursts occurring in the kilometric domain suggest that the shocks reach very close to L1 and hence are detected there. The narrow spectral range of type II bursts produced by the RLNS CMEs is thus consistent with the lower kinetic energy of the CMEs discussed in the previous paragraph. The SEP association rate is significant for the RLNS population (~50%), although much smaller compared to the RLS population (74%). The high degree of SEP association is consistent with the shock formation close to the Sun, irrespective of how the shocks evolved later on. This is further supported by the fact that most of the type II bursts (75% for RLNS and 72% for RLS) had components in the metric domain, indicating shock formation at heliocentric distances <2 Rs [*Gopalswamy* et al., 2009b; 2012].

The above discussion certainly points to the possibility that the shocks associated with the RLNS population generally die out before reaching 1 AU because of their low energy. However, when we look at the speed distributions in Fig. 4, we see that there are many CMEs in the RLS population that have low speeds suggesting that the kinetic energy may not be the only factor deciding the shock arrival at 1 AU. One of the obvious factors is the deflection by nearby coronal holes, which can make disk-center CMEs behave like limb CMEs in causing shock signatures at 1 AU [*Gopalswamy et al.*, 20009c]. In these events, the coronal holes deflected the CMEs such that the shock drivers did not reach Earth, but the shock flanks did. With a larger deflection and a weaker CME, even the shock may not arrive at Earth. In order to check this, we performed an analysis similar to what was done in *Gopalswamy et al.* [2009c]. We computed the coronal hole influence parameter (CHIP) defined as the average unipolar magnetic flux in the coronal hole divided by the square of the distance between the coronal-hole centroid and the CME source region. If more than one coronal hole is present, a vector sum of CHIP is computed, assuming that the vector points from the coronal-hole centroid to the source region. The CHIP values ranged from 0.16 to 10. *Gopalswamy et al.* [2009c] found that the average CHIP value for shocks that arrived at Earth along with the drivers was ~2.5. For the driverless shocks, the average value was 5.8. At least 5 CMEs in the RLNS population might have been subject to coronal-hole deflection because the CHIP exceeded 2. Since the CMEs in question are already weak, even a weaker CHIP might be able to push the shock away from the Sun-Earth line.

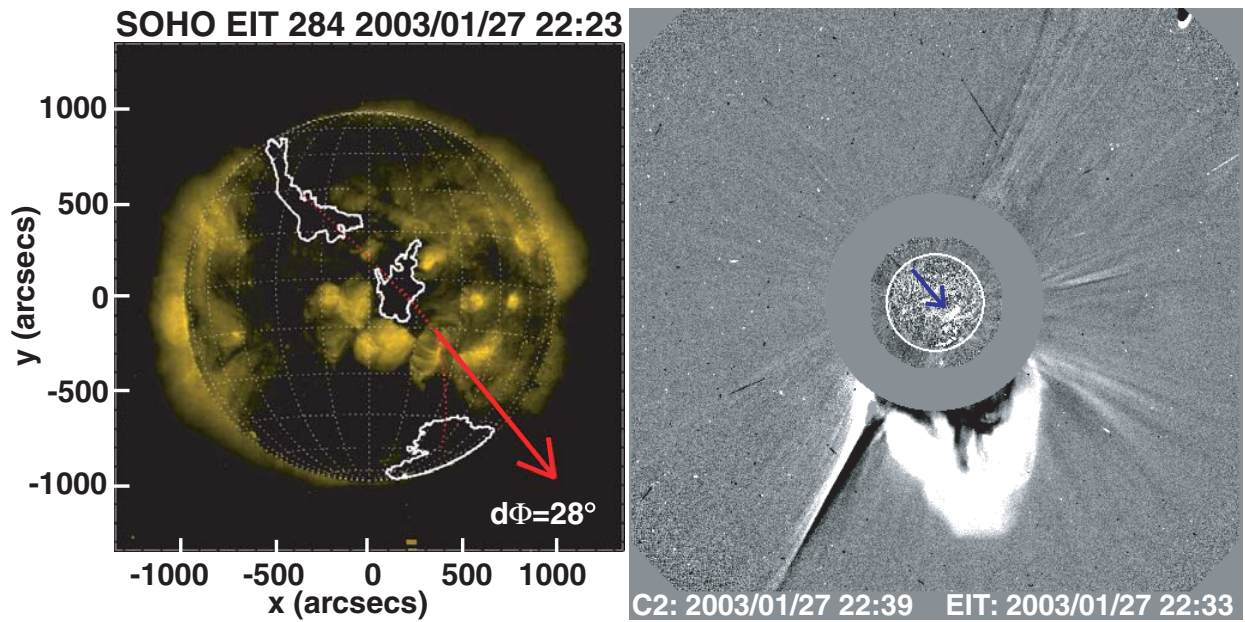

*Figure 10. The direction of the coronal hole influence (left) for the RLNS CME on 2003 January (right). The main contributor to the CHIP is the coronal hole near the disk center because the other two coronal holes had an order of magnitude smaller influence. The difference between the direction of the CME motion and the direction of CHIP, dΦ is 28º. The CME propagates generally to the South, even though the source region was at S17W23 (pointed by arrow in the LASCO image).*

Figure 10 shows one of the examples, where the CHIP was as high as 10 mainly due to a coronal hole that was situated between the disk center and the CME source region. Three coronal holes can be seen on the disk in the SOHO/EIT image taken around the time of eruption of the CME on 2003 January 27 at 22:33 UT. The CME originated from the southwest quadrant of the Sun (S17W23). However, the CME propagates mostly to the south. Since the coronal hole is located between the eruption region and the disk center, we expect the CME to be deflected away from the Sun-Earth line, which might have contributed to the lack of shock at 1 AU. The difference between the direction of CME propagation and the direction of CHIP is small (d$\Phi$ = 28°), which suggests that the coronal hole deflection is significant [*Gopalswamy et al.* 2009c]. The event with the next higher CHIP value (4.9) is the CME on 2004 April 8 at 13:30 UT. Two CMEs from this source region were reported to have driverless shocks because of coronal hole deflection on April 6 at 13:31 UT and April 8 at 10:30 UT [*Gopalswamy et al.*, 2009c]. The CHIP values were also large: 4.4 and 12, respectively. Both these CMEs were halos and their speeds exceeded 1000 km/s. The RLNS CME on April 8 at 13:30 UT had a similar speed (959 km/s), but was relatively narrow (the width was only 92°). Thus the coronal hole deflection is likely to have pushed the CME further away resulting in no shock at all. Note that we cannot rule out the possibility that the shocks died out before reaching 1 AU.

As a final point, we would like to mention that CME interaction is another reason as to why disk-center CMEs may not produce shock signature at 1 AU, as illustrated in Fig. 3. In the list of events we studied, there was only one such event. Such interactions are also known to result in complex ejecta at 1 AU [*Burlaga et al.*, 2002]. This generally happens when the preceding CME is slower than the following one. In addition to CME merger, it is also possible that CME trajectories can be changed when CMEs collide [*Gopalswamy et al.*, 2001a]. In the present list we do not have such events.

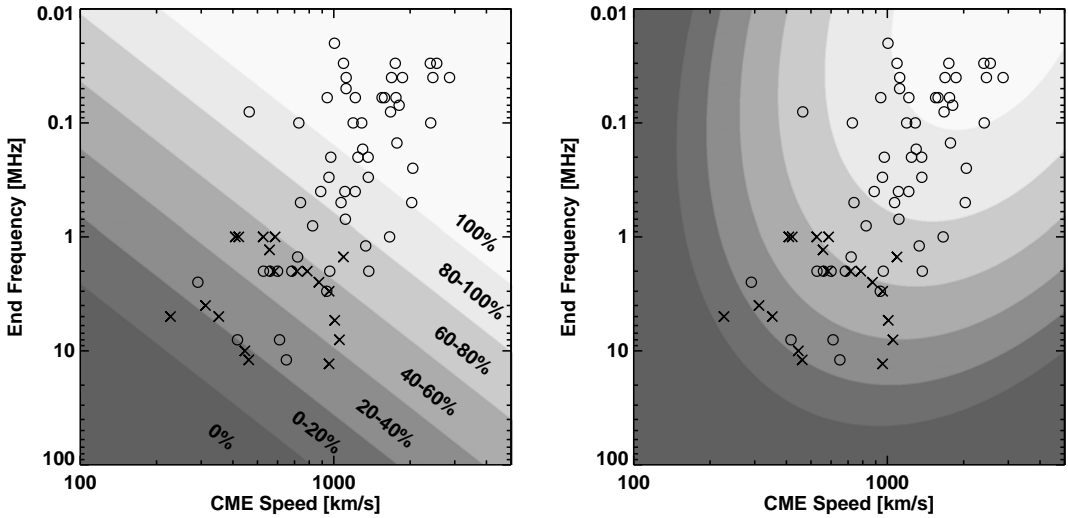

*Figure 11. 1-AU shock association rate of radio-loud CMEs originating from the disk center based on the 74 CMEs (53 RLS and 21 RLNS) shown as a function of the CME speed and the ending frequency of the associated type II burst. (a) contours based on linear extrapolation of the data points and (b) based on quadratic extrapolation using the IDL routine, "GRIDDATA". The contours of shock association rate are shown on the left plot. The darkest region corresponds to 0% probability and the lightest region corresponds to 100%. The actual data points are shown superposed (circles: RLS events; crosses: RLNS events).*

Looking at various reasons that a disk-center radio-loud CME may not produce a shock signature at 1 AU, we see that the CME speed and the ending frequency of type II bursts stand out as the primary factors. We can use these shock association rate with these two parameters as an indicator of the probability that a radio-loud CME results in a shock at 1 AU. This is shown Fig. 11 in a two-dimensional probability plot as a function of the CME speed and type II burst ending frequency (the contour intensity represents the probability). It is clear that the probability of detecting a shock at 1 AU increases rapidly when the CME speed exceeds ~1000 km/s and the type II bursts extends to frequencies below ~1 MHz. At lower speeds and higher type II ending frequencies, the probability falls below 60%. CMEs that do not produce a shock signature at 1 AU occupy the lower left corner of the plot. Clearly there is an overlap between RLS and RLNS events in this region. One may have to look into other possible reasons mentioned above to understand these events.

In summary, we can list the main findings of this investigation as follows:

1) A significant number of disk-center radio-loud CMEs did not produce shock signatures at Earth (~28%). Such RLNS CMEs were driving shocks near the Sun as evidenced by the associated type II radio bursts and SEP events.

2) The RLNS CMEs were generally slow, with the average speed (644 km/s) only slightly higher than that of the general population (~475 km/s), but fast enough to drive shocks near the Sun.

3) The fraction of halos in the RLNS CME population is much smaller than that in the RQS and RLS populations, suggesting a smaller width.

4) There is roughly equal number of accelerating and decelerating CMEs in the RLNS population unlike the RLS (mostly decelerating) and RQS (mostly accelerating) populations.

5) About half of the RLNS CMEs were associated with weak SEP events, suggesting that the presence of a shock near the Sun is important for accelerating SEPs.

6) The type II bursts in the RLNS population have a narrow spectral range with ending frequencies above ~1 MHz, which corresponds to a heliocentric distance of ~10 Rs. Beyond this distance, the shocks may be radio quiet or die out.

7) At least in a subset of the RLNS events, coronal hole deflection seems to be significant, which might have pushed the CMEs away from the Sun-Earth line so there is no shock signatures at Earth.

8) CME interaction is another process by which CMEs may not have shock at Earth.

9) The probability of shock association rapidly increases above 60% when the CME speeds exceed ~1000 km/s and when the type II ending frequency decreases below ~1 MHz.

**Acknowledgments:** We thank the SOHO, Wind and ACE science teams for making the shock data available on line. This research was supported by NASA LWS TR&T program. SOHO is a project of international cooperation between ESA and NASA.

**References**

Bougeret, J.-L., et al. (1995), Waves: The Radio and Plasma Wave Investigation on the Wind spacecraft, Space Sci. Rev., 71, 231.

Brueckner, G.E., et al. (1995), The large angle spectroscopic coronagraph (LASCO), Solar Phys., 162, 357.

Burlaga, L. F.; Plunkett, S. P.; St. Cyr, O. C. (2002), Successive CMEs and complex ejecta, J. Geophys. Res., 107, SSH 1-1, DOI 10.1029/2001JA000255

Cliver, E. W., S. W. Kahler, and D. V. Reames (2004), Coronal shocks and solar energetic proton events, Astrophys. J., 605, 902–910, doi:10.1086/382651.

Delaboudinière, J.-P., et al. (1995), EIT: The extreme-ultraviolet imaging telescope for the SOHO mission, Solar Phys., 162, 291.

Gopalswamy, N., (2004), Recent advances in the long-wavelength radio physics of the Sun, Planetary and Space Sci., 52, 1399-1413

Gopalswamy, N. (2006) Coronal Mass Ejections and Type II Radio Bursts, N. Gopalswamy in Solar Eruptions and Energetic Particles, Geophysics monograph 165, ed. N. Gopalswamy, R. Mewaldt, and J. Torsti, American Geophysical Union, Washington DC, p.207

Gopalswamy, N. ., A. Lara, R. P. Lepping, M. L. Kaiser, D. Berdichevsky, and O. C. St. Cyr (2000), Interplanetary acceleration of coronal mass ejections, Geophys. Res. Lett. 27, 145.

Gopalswamy, N., S. Yashiro, M. L. Kaiser, R. A. Howard and J.-L. Bougeret (2001a), Radio signatures of CME interaction: CME Cannibalism? Astrophys. J., 548, L91

Gopalswamy, A. Lara, M. L. Kaiser and J.-L. Bougeret (2001b), Near-Sun and Near-Earth Manifestations of Solar Eruptive Events, J. Geophys. Res., 106, 25,261

Gopalswamy, N., S. Yashiro, G. Michałek, M. L. Kaiser, R. A. Howard, D. V. Reames, R. Leske, and T. von Rosenvinge (2002), Interacting coronal mass ejections and solar energetic particles, Astrophys. J., 572, L103-L107

Gopalswamy, N., E. Aguilar-Rodriguez, S. Yashiro, S. Nunes, M. L. Kaiser, R. A. Howard (2005). Type II radio bursts and energetic solar eruptions, J. Geophys. Res., 110, A12S07, doi: 10.1029/2005JA011158

Gopalswamy, N., Yashiro, S., Akiyama, S. (2006), Coronal mass ejections and space weather due to extreme events, Proceedings of the ILWS Workshop. Goa, India. February 19-24, 2006. Editors: N. Gopalswamy and A. Bhattacharyya, Quest Publications, Mumbai, p.79.

Gopalswamy, N., S. Yashiro, S. Akiyama (2007), Geoeffectiveness of Halo Coronal Mass Ejections, J. Geophys. Res., 112, A06112, doi:10.1029/2006JA012149

Gopalswamy, N., S. Yashiro, H. Xie, S. Akiyama, E. Aguilar-Rodriguez, M. L. Kaiser, R. A. Howard, and J.-L. Bougeret (2008a), Radio-Quiet Fast and Wide Coronal Mass Ejections, Astrophys. J., 674, 560

Gopalswamy, N., S. Yashiro, S. Akiyama, P. Mäkelä, H. Xie, M. L. Kaiser, R. A. Howard and J. L. Bougeret (2008b), Coronal Mass Ejections, Type II Radio Bursts, and Solar Energetic Particle Events in the SOHO Era, Ann. Geophys., 26, 3033–3047, doi:10.5194/angeo-26-3033-2008.

Gopalswamy, N., S. Yashiro, G. Michalek, G. Stenborg, A. Vourlidas, S. Freeland, R. A. Howard (2009a), The SOHO/LASCO CME Catalog, Earth Moon and Planets, 104, 295-313

Gopalswamy, N., W. T. Thompson, J. M. Davila, M. L. Kaiser, S. Yashiro, P. Mäkelä, G. Michalek, J.-L. Bougeret, and R. A. Howard (2009b), Relation Between Type II Bursts and CMEs Inferred from STEREO Observations, Solar Phys., 259, 227-254

Gopalswamy, N., P. Mäkelä, H. Xie, S. Akiyama, S. Yashiro (2009c), CME interactions with coronal holes and their interplanetary consequences, J. Geophys. Res., 114, A00A22, DOI: 10.1029/2008JA013686

Gopalswamy, N., H. Xie, P. Mäkelä, S. Akiyama, S. Yashiro, M. L. Kaiser, R. A. Howard, andJ.-L. Bougeret (2010a), Interplanetary shocks lacking type II radio bursts, Astrophys. J., 710, 1111–1126, doi:10.1088/0004-637X/710/2/1111.

Gopalswamy, N., S. Yashiro, G. Michalek, H. Xie, P. Mäkelä, A. Vourlidas, and R. A. Howard (2010b), The catalog of halo coronal mass ejections from SOHO, Sun Geosphere, 5(1), 7–16.

Gopalswamy, N., Xie, H., Yashiro, S., Akiyama, S., Mäkelä, P., and Usoskin, I. G., Properties of Ground Level Enhancement Events and the Associated Solar Eruptions during Solar Cycle 23, Space Scie. Rev., in press, 2012, DOI: 10.1007/s11214-012-9890-4

Mäkelä, P., N. Gopalswamy, S. Akiyama, H. Xie, and S. Yashiro (2011), Energetic storm particle events in coronal mass ejection-driven shocks, J. Geophys. Res., 116, A08101, DOI: 10.1029/2011JA016683

Yashiro, S., N. Gopalswamy, G. Michalek, O. C. St. Cyr, S. P. Plunkett, N. B. Rich, and R. A. Howard (2004), A catalog of white light coronal mass ejections observed by the SOHO spacecraft, J. Geophys. Res., 109, A07105, doi:10.1029/2003JA010282.

Table 1. List of Disk-center CMEs associated with type II radio bursts with no shock at Earth

| Type II Start | End | F1 | F2 | m II | Location | AR | Imp | CME Time | CPA | W | V | SEP | CHIP |
|---|---|---|---|---|---|---|---|---|---|---|---|---|---|
| 1997/04/01 14:00 | 14:15 | 8 | 4 | Y | S25E16 | 8026 | M1.3 | 15:18 | 74 | 79 | 312 | Y | 1.7 |
| 1997/11/03 05:15 | 12:00 | 14 | 5 | Y | S18W16 | 8100 | C8.6 | 05:28 | 240 | 109 | 227 | N? | 0.27 |
| 1997/11/03 10:30 | 11:30 | 14 | 5 | Y | S16W21 | 8100 | M4.2 | 11:11 | 233 | 122 | 352 | Y | 0.29 |
| 1999/08/28 18:25 | 18:33 | 16 | 12 | Y | S26W14 | 8674 | X1.1 | 18:26 | 120 | 245 | 462 | Y | 2.4 |
| 2000/04/09 23:15 | 23:45 | 4.5 | 1 | Y | S14W01 | 8948 | M3.1 | 00:30[b] | Halo | 360 | 409 | HiB | 0.24 |
| 2000/09/25 02:20 | 03:00 | 14 | 1 | Y | N09W18 | 9169 | M1.8 | 02:50 | Halo | 360 | 587 | HiB | 0.58 |
| 2001/05/12 23:52 | 00:12[b] | 3 | 1 | Y | S17E02 | 9455 | M3.0 | 02:21[b] | 190 | 132 | 527 | N | 0.60 |
| 2001/09/16 13:50 | 14:10 | 3 | 2 | N | S13E15 | 9616 | C8.9 | 14:30 | 131 | 68 | 584 | HiB | 1.10 |
| 2001/09/17 08:35 | 08:47 | 14 | 5.4 | Y | S14E04 | 9616 | M1.5 | 08:54 | 198 | 166 | 1009 | HiB | 1.0 |
| 2001/09/20 18:43 | 18:57 | 14 | 10 | N | N09W11 | 9631 | M1.5 | 19:31 | 306 | 207 | 446 | HiB | 0.57 |
| 2001/10/19 01:15 | 02:25 | 14 | 1.3 | Y | N16W18 | 9661 | X1.6 | 01:27 | Halo | 360 | 558 | Y[e] | 1.60 |
| 2002/07/17 07:30 | 07:45 | 3.5 | 2 | Y | N21W17 | 10030 | M8.5 | 07:31 | 36 | 177 | 716 | HiB | 2.80 |
| 2002/12/19 21:45 | 22:30 | 14 | 1.5 | Y | N15W09 | 10229 | M2.7 | 22:06 | Halo | 360 | 1092 | Y[e] | 2.60 |
| 2003/01/27 22:20 | 22:26 | 11 | 8 | Y | S17W23 | FILA | C2.4 | 22:23 | 205 | 267 | 1053 | Y | 10.0 |
| 2004/04/08 13:30 | 14:00 | 6 | 3 | N | S19W08 | 10588 | C1.3 | 13:31 | 198 | 92 | 959 | Y | 4.90 |
| 2004/07/23 19:00 | 19:35 | 10 | 2.5 | N | S08W20 | 10652 | C1.6 | 19:31 | 209 | 100 | 874 | HiB | 0.95 |
| 2004/10/30 06:40 | 07:40 | 4 | 1 | Y | N13W22 | 10691 | M4.2 | 06:54 | Halo | 360 | 422 | Y[e] | 1.40 |
| 2005/01/17 09:25[a] | 09:50 | 14 | 0.4 | Y | N15W25 | 10720 | X2.2 | 09:30 | Halo | 360 | 2094 | Y | ---- |
| 2006/04/30 16:20 | 01:00[c] | 1 | 0.2 | Y | S09E08 | 10876 | C1.8 | 09:54 | Halo | 360 | 544 | N | 0.16[f] |
| 2006/08/26 20:40 | 21:00 | 7 | 2[d] | N | S10E08 | 10905 | C2.5 | 20:57 | 164 | 208 | 786 | N | 0.95 |
| 2007/05/19 13:02 | 13:05 | 16 | 13 | Y | N07W06 | 10956 | B9.5 | 13:24 | 260 | 106 | 958 | Y | 0.98 |

[a]CME interaction event; [b]Time corresponds to the day next to the one in column 1; [c]Time corresponds to two days after the one in column 1. [d]Partial WAVES data gap, so the lower frequency is uncertain. [e]minor GOES events. [f]This event may not be a DH event.

Table 2. Comparison of Properties of Disk-Center CMEs with and without shocks at 1 AU

| Property | RQS CMEs | RLS CMEs | RLNS CMEs |
| --- | --- | --- | --- |
| # of CMEs | 39 | 53 | 21[b] |
| Mean CME speed | 509 km/s | 1281 km/s | 644 km/s |
| Halo Fraction | 54% | 91% | 30% |
| Average acceleration | 7.5 m/s$^2$ | -13.6 m/s$^2$ | -1.9 m/s$^2$ |
| Flare Importance[a] | 0X 4M, 16C, 5B | 18X, 25M, 10C | 2X, 10M, 7C, 1B |
| SEPs (≥10 pfu) | None | 23 | None |
| SEPs (<10 pfu) | 2? | 14 | 10 |

[a]14 events did not have flare data, [b]The 2005 January 17 event at 09:25 UT has been excluded in determining the properties because it is an interaction event